\documentclass[aps,prl,twocolumn,groupedaddress]{revtex4}

\usepackage{graphics}
\usepackage{epstopdf}
\usepackage{graphicx}% Include figure files
\usepackage{dcolumn}% Align table columns on decimal point
\usepackage{bm}% bold math

\begin{document}

%Title of paper
\title{Quantum oscillations of dissipative resistance in crossed electric and magnetic fields}

\author{Scott Dietrich}
\author{Sean Byrnes}
\author{Sergey Vitkalov}
\email[Corresponding author: ]{vitkalov@sci.ccny.cuny.edu}
%\homepage[]{Your web page}
%\thanks{}
%\altaffiliation{}
\affiliation{Physics Department, City College of the City University of New York, New York 10031, USA}
\author{D. V. Dmitriev}
\author {A. A. Bykov}
\altaffiliation{Novosibirsk State Technical University, 630092 Novosibirsk, Russia}
\affiliation{Institute of Semiconductor Physics, 630090 Novosibirsk, Russia}

\date{\today}

\begin{abstract} 
Oscillations of dissipative resistance of two-dimensional electrons in GaAs quantum wells are observed in response to an electric current $I$ and a strong magnetic field applied perpendicular to the two-dimensional systems. Period of the current-induced oscillations does not depend on the magnetic field and  temperature. At a fixed current the oscillations are periodic in inverse magnetic fields  with a period that does not depend on dc bias. The proposed model considers  spatial variations of electron filling factor, which are induced by the electric current,  as the origin of the resistance oscillations.   
\end{abstract}
  

\maketitle

\section{I. Introduction}

Nonlinear transport properties of two-dimensional electrons, placed in quantizing magnetic fields, attract a great deal of attention both for its fundamental importance and remarkable properties found in highly mobile electron systems. In response to both microwave radiation and dc excitations, strongly nonlinear electron transport \cite{zudov2001,engel2001,yang2002,dorozh2003,willett2004,mani2004,kukushkin2004,stud2005,bykov2005,bykovJETP2006,bykov2007R,zudov2007R,du2007,stud2007,zudovPRB2008,bykov2008,romero2008,gusev2008a,zudov2009,hatke2009a,dorozh2009,vitkalov2009,vitkalov_review2009,gusev2009,durst2003,ryzhii1970,anderson,shi,liu2005,dietel2005,inarreaPRB2005,vavilov2004,dmitriev2005,alicea2005,volkov2007,glazman2007,dmitriev2007} that gives rise to unusual electron states \cite{mani2002,zudov2003,zudov2007,bykov2007zdr,zudov2008zdr,bykov2010,zudov2010,gusev2011,andreev2003,auerbach2005} has been reported and investigated. Very recent experimental studies in the low frequency domain \cite{bykov2007R,vitkalov2009,gusev2009,vitkalov_review2009} reveal that the dominant mechanism of the nonlinearity is related to a peculiar quantal heating (``inelastic" mechanism \cite{dmitriev2005}), which may not increase the broadening of electron distribution (``temperature") in systems with discrete spectrum \cite{vitkalov2009,vitkalov_review2009}. Due to this extraordinary property, the Joule heating strongly affects  the electron transport in quantum conductors. Microwave studies of the nonlinearity \cite{hatke2009a} in very high mobility systems  indicate the relevance of another nonlinear mechanism:  electric field induced variations in the kinematics of electron scattering on impurities  (``displacement" mechanism \cite{ryzhii1970,durst2003,vavilov2004}), which limits the lifetime of  an electron in a quantum state. The interplay between these two mechanisms has been investigated theoretically \cite{glazman2007}.

In this paper we show that at higher magnetic fields there is an additional  nonlinear  mechanism, which  induces substantial oscillations of the electron resistance in response to the applied electric current.  The period of the oscillations does not depend on the magnetic field. The oscillations are observed at low temperatures and strong magnetic fields, at which the quantum [Shubnikov-de Haas (SdH)] oscillations are well developed. The current-induced oscillations correlate with the SdH oscillations and are periodic in inverse magnetic fields. The oscillations are absent at smaller magnetic fields at which the SdH oscillations are also small or absent. The oscillations are found in samples with long quantum electron lifetime $\tau_q$ = 4 (ps) and are not observed in systems with broad Landau levels [$\tau_q$ = 1 (ps)].  

The proposed theoretical model considers the oscillations as a result of the electrostatic redistribution of the electron density, which induces the electric field and, thus, the electric current in the systems. The electron redistribution occurs across the sample and is associated with a spatial variation of the number of occupied Landau levels.  The model indicates that the resistance oscillates with the electric current with a period that does not depend on the magnetic field and the temperature.
 
\section{II. Experimental Setup}

Our samples are high-mobility GaAs quantum wells grown by molecular beam epitaxy on semi-insulating (001) GaAs substrates. The width of the GaAs quantum well is 13 nm. Two AlAs/GaAs type-II superlattices grown on both sides of the well served as barriers, providing a high mobility of two-dimensional (2D) electrons inside the well at a high electron density\cite{fried1996}. This is an important property of our samples and is discussed below in more detail.
Two samples (N1, N2) were studied with electron density $n_{1,2}$ = 8.2 $\times 10^{15}$ (m$^{-2}$), mobility $\mu_{1,2}$ = 93 (m$^2$/Vs)  and quantum lifetime $\tau_q = 4$ (ps).  Another two samples (N3, N4) had similar electron density $n_3$ = 8.2 $\times 10^{15}$ (m$^{-2}$), $n_4$ = 12.2 $\times 10^{15}$ (m$^{-2}$),  and mobility $\mu_3$ = 86 (m$^2$/Vs), $\mu_4$ = 89 (m$^2$/Vs), but much shorter quantum lifetime $\tau_q = $ 1 (ps).  

\begin{figure*}[t!]
\includegraphics[width=6in]{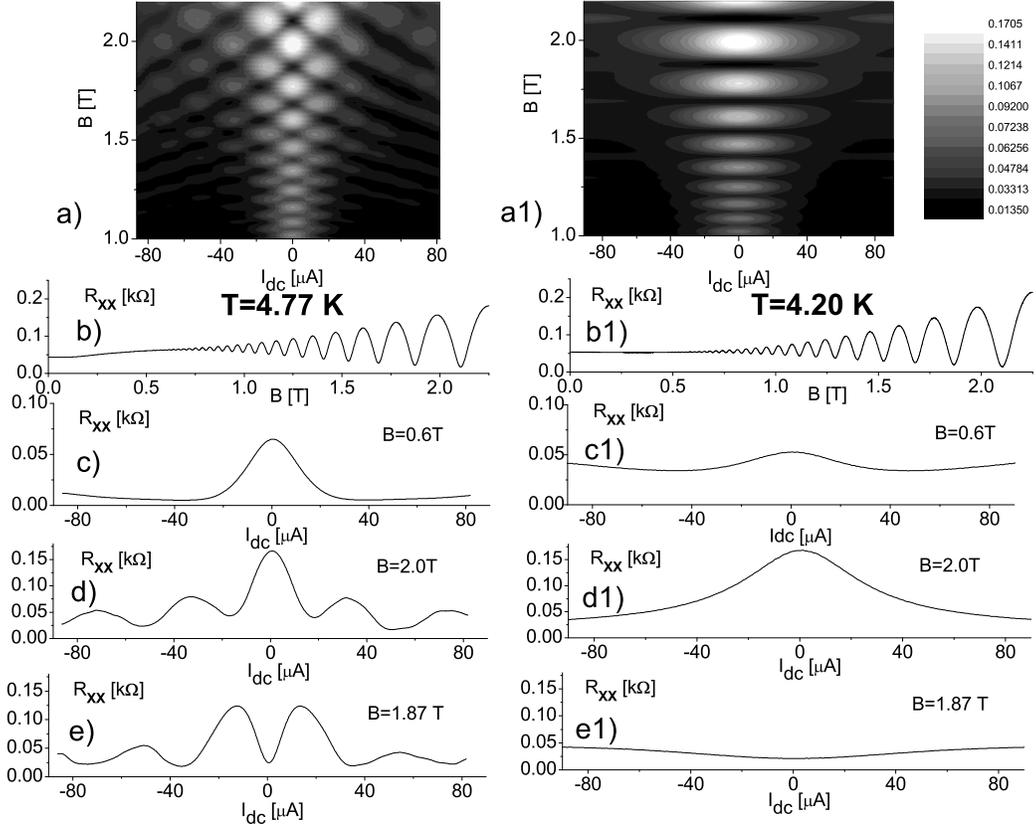}
\caption{Dependence of dissipative differential resistance $R_{xx}$ on magnetic field $B$ and electric current $I_{dc}$. Left panel presents data for sample N1 with quantum scattering time $\tau_q$ = 4 (ps) at temperature $T$ = 4.77 (K):  (a) 2D plot $R_{xx}$ vs $B$ and $I_{dc}$; (b) $R_{xx}$ vs $B$ at $I_{dc}$ = 0 ($\mu$A); (c) $R_{xx}$ vs $I_{dc}$ at $B$ = 0.6 (T);  (d) $R_{xx}$ vs $I_{dc}$ at $B$ = 2.0 (T); (e) $R_{xx}$ vs $I_{dc}$ at $B$ = 1.87 (T). Right panel presents data for sample N3 with quantum scattering time  $\tau_q$ = 1 (ps) at temperature $T$ = 4.2 (K):  (a1) 2D plot $R_{xx}$ vs $B$ and $I_{dc}$; (b1) $R_{xx}$ vs $B$ at $I_{dc}$ = 0 ($\mu$A); (c1) $R_{xx}$ vs $I_{dc}$ at $B$ = 0.6 (T);  (d1) $R_{xx}$ vs $I_{dc}$ at $B$ = 2.0 (T); (e1) $R_{xx}$ vs $I_{dc}$ at $B$ = 1.87 (T). }
\label{twosamples}
\end{figure*}

The studied 2D electron systems are etched in the shape of a Hall bar. The width and the length of the measured part of the samples are d = 50$\mu m$ and L = 250$\mu m$. To measure the resistance we have used the four probes method. Direct electric current $I_{dc}$ ($dc$ bias) is applied simultaneously with 12 Hz ac excitation $I_{ac}$ through the same current contacts ($x$ direction).
The longitudinal and ac (dc)  voltage $V^{ac}_{xx}$ ($V^{dc}_{xx}$) is measured between potential contacts displaced 250$\mu m$ along each side of the sample. The Hall voltage $V_H$ is measured between potential contacts displaced 50$\mu m$ across the electric current in $y$ direction.

The current contacts are sufficiently separated from the measured area by a distance of 500$\mu m$, which is much greater than the inelastic relaxation length of the 2D electrons $L_{in}=(D \tau_{in})^{1/2} \sim 1 - 5 $ ($\mu m$). This ensures that the current contacts do not affect the results of the measurements.  The longitudinal and Hall $ac$ voltages were measured simultaneously using two lockin amplifiers with 10-M$\Omega$ input impedances. dc voltages were measured, using high impedance (1 G$\Omega$) voltmeters. The potential contacts provided insignificant contribution to the overall response due to small values of the contact resistance (about 1K$\Omega$) and negligibly small electric current flowing through the contacts.

Measurements were carried out for different temperatures in the range of 0.3-10 Kelvin in a He-3 insert in a superconducting solenoid. Samples and a calibrated thermometer were mounted on a cold copper finger in vacuum. Magnetic fields were applied perpendicular to the 2D electron layers.

\section{III. Results}

\begin{figure*}[t!]
\includegraphics[width=6in]{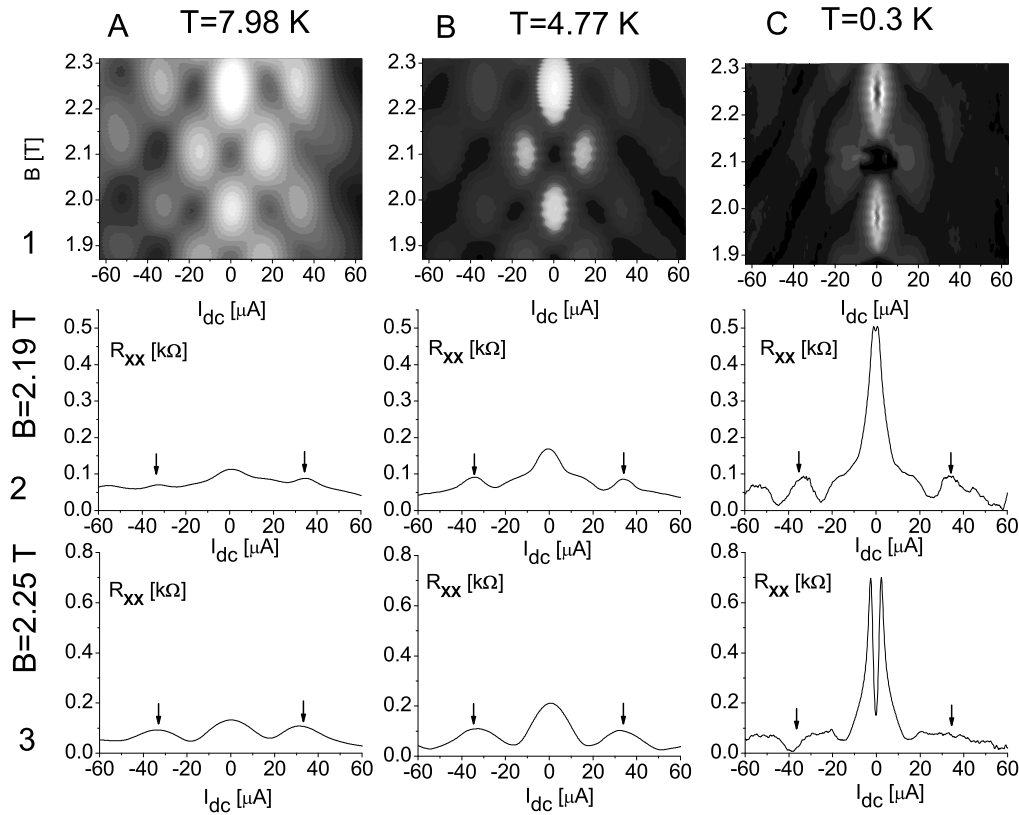}
\caption{Panels (a1), (b1) and (c1) present dependence of differential resistance of sample N1 on magnetic field and dc bias taken at different temperatures as labeled. Bright(dark) spots indicate high(low) resistance. Panels (a2), (b2), and (c2)  present horizontal cuts of the plots shown in (a1), (b1), and (c1) at magnetic field $B$ = 2.19 (T). Figures (a3), (b3), and (c3) present horizontal cuts of the plots shown in(a1), (b1), anb (c1) taken at magnetic field $B$ = 2.25 (T).}
\label{temperature}
\end{figure*}

Figure {\ref{twosamples}} presents the magnetoresitance of two 2D electron systems with approximately the same electron density but with different electron lifetime $\tau_q$. The left panel, Figs. {\ref{twosamples}}(a)-{\ref{twosamples}}(e), shows data taken at temperature $T$ = 4.77 K for sample N1 with $\tau_q$ = 4 (ps). Figure {\ref{twosamples}}(a) demonstrates an overall behavior of the differential resistance at different dc currents from -80 to 80 ($\mu$A) and magnetic fields from 1 to 2.25 T. Taken at zero dc bias [$I_{dc}$ = 0 ($\mu$A)] vertical cut of the 2D plot corresponds to the linear response of the system. The cut, extended to zero magnetic field, is shown in Fig. {\ref{twosamples}}(b). The figure demonstrates well-known Shubnikov de Haas (SdH) oscillations of the resistance. These oscillations are periodic in the inverse magnetic field $1/B$. Figure {\ref{twosamples}}(c) demonstrates a horizontal cut of the 2D plot, which is taken at magnetic field $B$ = 0.6 T.  At this magnetic field the SdH oscillations are absent as shown in Fig. {\ref{twosamples}}(b). The strong decrease of the resistance with the dc bias is due to quantal heating, which is studied for this sample in detail inRef. \cite{vitkalov2009} (see also Ref. \cite{gusev2009} ). Figure {\ref{twosamples}}(d) presents another dependence of the resistance on the dc bias. The dependence is taken at a maximum of SdH oscillations and corresponds to a horizontal cut of the 2D plot at B = 2 T. Figure {\ref{twosamples}}(d) shows oscillations of the resistance with the dc bias. Figure {\ref{twosamples}}(e) shows a dc bias dependence of the resistance taken at minimum of SdH oscillations at B = 1.87 T. Figure {\ref{twosamples}}(e) demonstrates oscillations, which are complementary to the oscillations shown in Fig. {\ref{twosamples}}(d). Sample N2 exhibits similar oscillations (not shown). The oscillations presented in Figs.  {\ref{twosamples}}(a),{\ref{twosamples}}(d) and {\ref{twosamples}}(e)  are the main subject of this paper \cite{stud2011}.

The right panel of Fig. {\ref{twosamples}} presents data obtained for sample N3 with similar electron density but with considerably shorter quantum scattering time $\tau_q$ = 1 (ps). The data are taken at temperature $T$ = 4.2 K. Figures {\ref{twosamples}}(a1)-{\ref{twosamples}}(e1) demonstrate dependencies taken at the same conditions as the dependencies presented in Figures {\ref{twosamples}}(a)-{\ref{twosamples}}(e). Due to the shorter time $\tau_q$ the Landau levels in the sample N3 are considerably broader than the quantum levels in the sample N1 and overlap substantially at $B$ = 0.6 (T) (see Fig.2 in Ref.\cite{vitkalov2009} ). In result shown in Fig. {\ref{twosamples}}(c1) resistance variations  are considerably smaller the one shown in Fig.{\ref{twosamples}}( c) \cite{dmitriev2005,vitkalov2009}.  Figures {\ref{twosamples}}(a1), {\ref{twosamples}}(d1) and {\ref{twosamples}}(e1) exhibit qualitatively different behavior: sample N3 does not show any oscillations with the dc bias.

Figure {\ref{temperature}} demonstrates the effect of temperature on these oscillations. Figures {\ref{temperature}}(a1), {\ref{temperature}}(b1) and {\ref{temperature}}(c1) present the dependence of the differential resistance on magnetic field and dc bias taken at different temperatures as shown. The amplitude and shape of the oscillations depend on the temperature but the positions of the oscillations are essentially the same at different temperatures. At the lowest temperature $T$ = 0.3 (K) spin splitting of the Landau levels is observed. The splitting makes correlations between different curves less obvious.  Figures {\ref{temperature}}(a2), {\ref{temperature}}(b2) and {\ref{temperature}}(c2) present horizontal cuts of the corresponding 2D plots {\ref{temperature}}(a1), {\ref{temperature}}(b1) and {\ref{temperature}}(c1) taken at magnetic field $B$ = 2.19 Tesla. At this magnetic field a maximum of SdH oscillations, which correspond to a spin polarized Landau level, is observed at $T$ = 0.3 (K).  The cuts indicate maximums  at around +35 and -35 ($\mu$A) for all three temperatures. Arrows mark the maximums. The magnitude of the oscillations increases as the temperature decrease.  Figures {\ref{temperature}}(a3), {\ref{temperature}}(b3) and {\ref{temperature}}(c3) present horizontal cuts taken at magnetic field B = 2.25 (T). These cuts correspond to a SdH resistance maximum at high temperatures, which evolves into a minimum at lowest temperature $T$ = 0.3 (K) at which the spin splitting is larger the temperature.  Two cuts taken at T = 7.98 (K) and at the T = 4.75 (K) [Figs. {\ref{temperature}}(a3) and {\ref{temperature}}(b3)] demonstrate good correlation. At lowest temperature the resistance demonstrates minimum at zero dc bias and maximums at +35 and -35 ($\mu$A) are not observed. 

\begin{figure}[t!]
\includegraphics[width=3.5in]{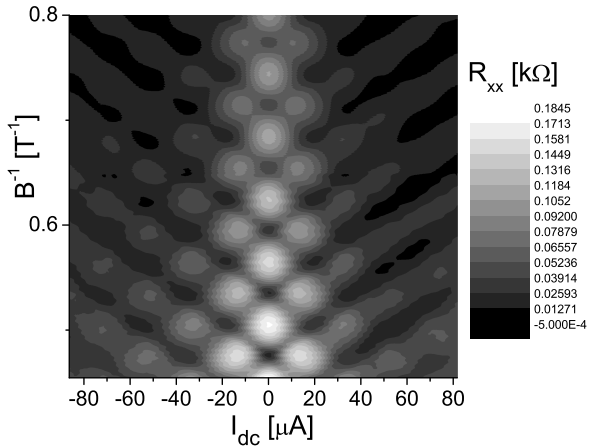}
\caption{Dependence of the dissipative resistance on inverse magnetic field and dc bias. T=4.77 (K). Sample N1.
}
\label{position}
\end{figure}

Figure {\ref{position}} presents dependence of the differential resistance on the inverse magnetic field and dc  bias for sample N1. The plot emphasizes the periodicity of the observed oscillations with respect to both the dc bias and the inverse magnetic field $1/B$. The figure indicates that the positions of the oscillations with respect to the dc bias do not change considerably with almost two times variation of the magnetic field.   

\begin{figure}[t!]
\includegraphics[width=3.5in]{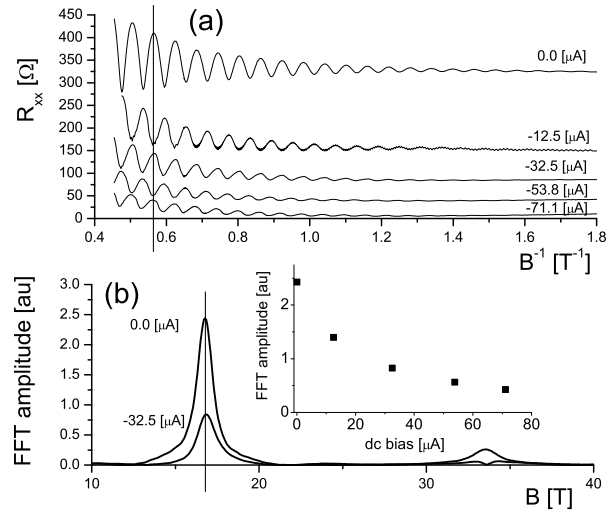}
\caption{(a)Dependence of the dissipative resistance on inverse magnetic field at different dc biases as labeled. The curves are shifted from the top to the bottom by 260, 120, 80, 35 and 0 ($\Omega$) for clarity; (b) Fourier transformation of the oscillations shown in (a) at two dc biases as labeled. Inset shows dependence of the first harmonic of the oscillations on dc bias. T=4.77 (K). Sample N1.
}
\label{fft}
\end{figure}

Figure \ref{fft}(a) presents vertical cuts of Fig.\ref{position} taken at different dc biases, which are close to maximums and minimums shown on Fig.\ref{twosamples}(d). The figure demonstrates that the $1/B$ periodic oscillations at $I_{dc}$ = -32.5 and -71.1 $\mu$A are in phase whereas the oscillations at $I_{dc}$ = -12.5 and -53.8 $\mu$A are 180$^\circ$ shifted with respect to SdH oscillations at zero dc bias. Figure \ref{fft}(b) presents a Fourier spectrum of the oscillations at $I_{dc}$ = 0 and -32.5 $\mu$A. The inset shows a dependence of the amplitude of the first harmonic of the oscillations on the dc bias. The experiment indicates a reduction of the oscillations with the dc bias increase.

Figures \ref{position} and \ref{fft}  demonstrates strong correlation of the dc biased-induced oscillations with the quantum oscillations at zero dc bias (SdH oscillations). This is an indication that these oscillations have a common origin. Below we consider a model, in which the oscillations are induced by spatial variations of the number of occupied Landau levels across the Hall bar sample. 

\section{IV. Model and Discussion}

Shubnikov-de Haas oscillations occur due to quantization of electron spectrum in a magnetic field \cite{shoenberg1984}. With an increase of the magnetic field, energy gaps between Landau levels increase and the top occupied Landau level intersects the Fermi energy $E_F$. At this condition resistivity of the electron systems is at a maximum. When the Fermi energy is between two Landau levels the resistivity is at a minimum. Thus the resistance oscillates with variations of the number of the Landau levels occupied by electrons.

We propose that the dc bias-induced oscillations also occur due to a variation of the electron filling factor but, in contrast to SdH oscillations, the variation appears across the sample and is related to a spatial change of electron density $\delta n$.  If the change is comparable with the number of electron states in a Landau level $n_0=m/(\pi \hbar^2) \cdot \hbar \omega_c$, then one should expect a variation of the electron resistivity. As shown below the spatial variation of the resitivity leads to  oscillations of the sample resistance. 

A simple electrostatic estimation demonstrates that in a vacuum the variation of electron density $\delta n \sim n_0$ creates a voltage, which is on several orders of magnitude stronger than the one observed in the experiment. The estimation dictates, therefore, the presence of a strong screening of electric charges $e\delta n$ in the samples. The proposed model assumes that the screening is due to Xelectrons, which are located near the conducting 2D layer. 

\begin{figure}[t!]
\includegraphics[width=3.5in]{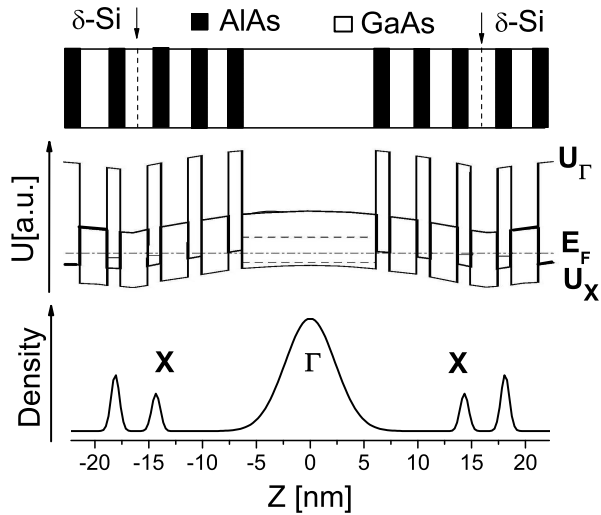}
\caption{Schematic diagram of a GaAs quantum well with
AlAs/GaAs short-period superlattice barriers. The two
lower plots show the Fermi energy level $E_F$, the edges of the
conduction band $U_{\Gamma}$  and $U_X$ and the density distributions of
$\Gamma$ and $X$ electrons.
}
\label{sample}
\end{figure}

Figure \ref{sample} shows a schematic diagram of our samples. The conducting GaAs quantum well is sandwiched between two layers of AlAs/GaAs superlattices (SLs) of the second kind \cite{fried1996}. The main purpose of the X electrons is to enhance  the electron mobility by screening the charged impurities near the conducting 2D layer.     The parameters of the superlattices are adjusted to set the system close to a metal-insulator transition. At this condition the barely conducting SL layers efficiently screen  electric charges and do not contribute considerably to the overall conductivity of the structure.

To estimate parameters relevant to the screening of the electron density $\delta n$ we consider the superlattice as a  metallic sheet placed at a distance $d$ from the conducting layer. A spatial variation of the electron density  $\delta n$ induces a variation of the voltage $V(r)$ across the layer: $e\delta n(r)= C V(r)$, where $C=\epsilon \epsilon_0/d$ is capacitance of the structure per unit area, $\epsilon=12$ is lattice permittivity, and $\epsilon_0=$ is permittivity of free space. A typical electric potential in the present experiments is $V$ = 60 mV at $B$ = 2 (T). This yields $d \sim \epsilon \epsilon_0 V/(en_0)$ = 39 (nm). This distance is comparable with  the thickness of the superlattice: 27-80 (nm).

Electric contacts connect the GaAs and the SL layers. Thus the system is considered as a set of parallel conductors. At zero magnetic field the distribution of the electric potential driving the current  is the same in all layers due to the same shape of the conductors. That is to say at B = 0 the potential difference between different layers is absent. In the poorly conducting SL layers the electric current is  several order of magnitude smaller than the one in the highly conducting GaAs quantum well.

\begin{figure}[t!]
\vskip -2.5 cm \hskip -2.8 cm 
\includegraphics[width=4 in]{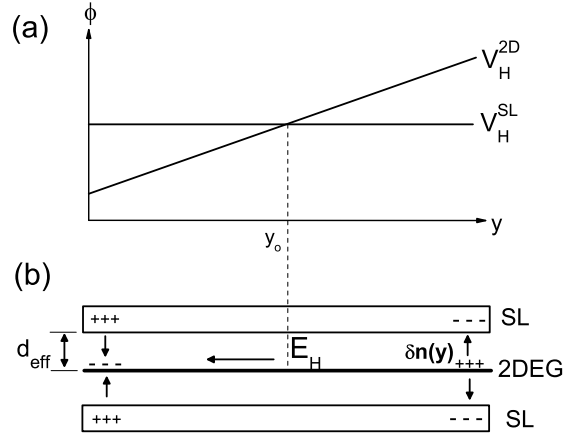}
\caption{ Dependence of the electric potential on position $y$ in the direction perpendicular to the electric current in strong magnetic field. Line $V_H^{2D}$ describes the potential in GaAs quantum well, in which strong Hall effect is developed. Line $V_H^{SL}$ describes the potential in the highly resistive superlattice layer, in which the Hall voltage is negligibly small due to the negligibly small current in the layer.  }
\label{sample2}
\end{figure}

The layers have a different distribution of the electric potential in a strong magnetic field, at which $\omega_c \tau_{tr}^{2D} \gg 1$ and $\omega_c \tau_{tr}^{SL} \ll 1$, where $\tau_{tr}^{2D}$ and $\tau_{tr}^{SL}$ are transport times in the GaAs and in the SL layers.  The distribution is shown in Fig. \ref{sample2}(a) for a small total current (linear response). At $\omega \tau_{tr}^{2D} \gg 1$ the electric field in the GaAs layer is almost perpendicular to the current due to the strong Hall effect.  In contrast, the very small electric current in the SL layer induces a Hall voltage, which is negligible.  The Hall voltages are shown in the Fig. \ref{sample2}(a). Figure \ref{sample2}(b) presents distribution of electric charges in the structure. Electric charges are accumulated near the edges of the 2D highly conducting GaAs layer, inducing the Hall electric field $E_H$. The charges are partially screened by charges accumulated in the conducting SL layers. 

Due to the small Hall voltage $E_H^{SL}$ and the absence of the electric current across the system the change of the electric potential $\phi^{SL}(y)$ in the SL  layer is negligibly small. Below we consider the potential $\phi^{SL}$ as a constant. Due to a finite screening length $\lambda_s$ in the SL layer the charge accumulation occurs at a distance $d \sim \lambda_s$. Below we approximate the charge distribution by a charged capacitor with an effective distance $d_{eff}$ between conducting plates.

A simplified model of the observed oscillations is presented below.  The model considers a long 2D Hall bar with a width $L_y$\cite{shashkin1986,dyakonov1990}. Electric current is in the $x$ direction and the Hall electric field is in the $y$ direction. In  a long conductor the electric field $\vec E=(E_x, E_y)$ is independent on $x$, due to the uniformity of the system in $x$ direction:  
\begin{equation}
\frac{\partial E_x}{\partial x} = \frac{\partial E_y}{\partial x}=0
\label{long}
\end{equation}

For a steady current Maxwell equations yield:

\begin{equation}
\frac{\partial E_x}{\partial y} = \frac{\partial E_y}{\partial x}
\label{max}
\end{equation}

Equations (\ref{long}) and eq.({\ref{max}}) indicate, that the $x$ component of the  electric field is the same at any location: $E_x=E=$const. 

Boundary conditions and the continuity equation require that the density of the electric current in $y$ direction is zero: $J_y=0$ and therefore,

\begin{equation}
E_x=\rho_{xx}J_x \hskip 0.5 cm E_y=\rho_{yx} J_x 
\label{E}
\end{equation} 
where $\rho_{xx}$ and $\rho_{yx}$ are longitudinal and Hall components of the resistivity tensor \cite{ziman}.  We approximate the SdH oscillations of the resistivity by a simple expression \cite{ando}:

\begin{equation}
\rho_{xx}(n(y))=\rho_D\left [1-\alpha \cdot cos\left(\frac{2\pi n}{n_0}\right) \right ] 
\label{sdh}
\end{equation}   
where $\rho_D$ is Drude resistivity and $\alpha$ describes the amplitude of the quantum oscillations. At a SdH maximum (minimum) filling factor $\nu=n/n_0$ is half integer (integer).

An electrostatic evaluation of the voltage between conducting layers, shown in Fig.\ref{sample2}(b), yields:

\begin{equation}
\phi^{2D}(y)=\phi^{SL}+\frac{e\delta n(y)d_{eff}}{2\epsilon \epsilon_0}
\label{voltage}
\end{equation}
where $\phi^{2D}$ and $\phi^{SL}$ are electric potentials of the GaAs [2D electrong as (2DEG)] and superlattice (SL) layers, and $\epsilon$ is permittivity of the SL layer. 
Expressing the electron density $\delta n$ in terms of electric potential $\phi^{2D}$ from Eq.{\ref{voltage}} and  substituting the relation into Eq.({\ref{sdh}}) one can find dependence of the resistitivity on the electric potential: $\rho_{xx}(\phi^{2D})$. 

The relation $E_y=-d\phi^{2D}/dy$ together with  Eq.(\ref{E}) yields:
\begin{equation}
-\frac{d\phi^{2D}}{dy} \rho_{xx}(\phi^{2D})=\rho_{yx}E
\label{main}
\end{equation}          

Separation of the variables $\phi^{2D}$ and $y$ and subsequent integration of Eq.({\ref{main}) between two sides of the 2D conductor ($y$ direction) with corresponding electric potentials $\phi_1$ and $\phi_2$ yield the following result: 

\begin{equation}
\matrix{ \rho_{D} ( \phi_2-\phi_1-\frac{\alpha}{\beta}\{sin[\beta (\phi_2-\phi_1)] \cr 
\times cos[\beta(\phi_2+\phi_1)+\theta_0]\} )=  \rho_{xy}EL_y \cr 
\cr
\beta=2\pi \epsilon_0 \epsilon/(e d_{eff} n_0), \cr
\theta_0=2\pi n/n_0-2\beta \phi^{SL}
}
\label{pr1}
\end{equation}
where $L_y$ is a width of the sample. Taking into account that longitudinal voltage is $V_{xx}=EL_x$, where $L_x$ is a distance between the potential contacts, and the Hall voltage $V_H=\phi_2-\phi_1= -\int E_y dy= -\rho_{yx}I$ [see Eq.(\ref{E})], the following relation is obtained:

\begin{equation}
V_{xx}=R_{D}\left (I-\frac{\alpha}{\beta \rho_{xy}}{sin(\beta\rho_{xy}I)\cdot cos[\beta (\phi_2+\phi_1)+\theta_0]}\right)
\label{final}
\end{equation}
, where $R_D=L_x\rho_D/L_y$ is Drude resistance. 

Equation \ref{final} is  simplified further for filling factors  corresponding to a minimum or a maximum of SdH oscillations. In this case the voltage  $\phi^{2D}(\delta y)-\phi^{SL}$ is expected to be an asymmetric function of the relative position $\delta y=y-y_0$ with respect to the center of the sample $y_0$   : $\phi_1-\phi^{SL} =-(\phi_2-\phi^{SL})$.\cite{pikus1992}  An example of the asymmetric distribution of the electric potential  is shown in Fig. \ref{sample2}(a) for small currents. In this case $\phi_1+\phi_2= 2\phi^{SL} $ and the argument of the cosine in Eq.(\ref{final}) becomes independent on the electric current. For the integer (a SdH minimum) and half-integer (a SdH maximum) filling factors the differential resistance $r_{xx}=dV_{xx}/dI$ is found to be

\begin{equation}
r_{xx}=R_D\left[1-\alpha \cdot cos\left(2\pi\frac{I}{I_0}\right)\cdot cos\left(2 \pi \frac{n}{n_0}\right)\right]
\label{final2}
\end{equation}    
where $I_0=\frac{e^2}{\pi \hbar} \frac{e d_{eff} n}{\epsilon \epsilon_0}$.

Equation \ref{final2} demonstrates periodic oscillations of the differential resistance with both the electric current $I$  and the inverse magnetic field  ($n_0 \sim B$). The period of the current induced oscillations $I_0$ does not depend on the magnetic field and temperature in accordance with the experiment. The phase difference between oscillations starting at the SdH maxima and minima is $\pi$, which is  in agreement with Figs.\ref{twosamples}(d) and \ref{twosamples}(e). The period of the oscillations shown above  $I_0 \approx 35$ ($\mu$A) indicates that the screening occurs at an effective distance $d_{eff} \approx$ 36 (nm). The distance is comparable with  thicknesses of the SL layers: 27 and 76 nm. 

The 1/B periodic oscillations of the resistance $R_{xx}$ at $I=i \cdot I_0$ (i=1,2...)  are in-phase  with SdH oscillations [I=0 (A)] whereas a phase of the oscillations at $I=(i-1/2) \cdot I_0$ is shifted by $\pi$ with respect to the phase of the SdH oscillations. This is in agreement with the results presented in Fig.\ref{fft}(a).

Figures \ref{twosamples}(d), \ref{twosamples}(e), and \ref{fft}(b) show that the amplitude of the oscillations depends on the current: the oscillations are weaker at a higher current. This behavior is beyond the simplified model presented above. 
There are  several possible mechanisms which may affect the amplitude of the quantum oscillations. 
One of the possibilities is the Joule heating. The heating may significantly decrease  the amplitude of quantum oscillations \cite{romero2008,vitkalov2009} reducing the magnitude of the spatial variations of the local resistivity. 

The current induced oscillations are absent in samples N3 and N4. These samples have the same electron densities and mobilities  as  samples N1 and N2 but four times shorter quantum scattering time $\tau_q$. We suggest that the observed significant difference in the $\tau_q$ and the absence of the oscillations  is result of a less effective screening in the SL layers of the samples N3, N4. A weaker screening is expected in  conducting superlattices, which are closer to the metal-insulator transition. In this case the screening of an electric charge occurs at a larger distance $\lambda_s$ due to  smaller density of conducting states. Thus the effective thickness $d_{eff} \sim \lambda_s$ and therefore the period $I_0$ can be significantly larger in weaker conducting SL layers.

\section{V. Conclusion}

Oscillations of differential resistance are observed in response to both electric current and magnetic field, which is  applied perpendicular to 2D electrons in GaAs quantum wells. The oscillations are periodic with the current and with the inverse magnetic field. The period of the current induced oscillations does not depend on magnetic field and temperature.  The SdH oscillations are a part of the set at zero dc bias. The proposed model considers spatial variations of the electron filling factor, which are induced by applied dc bias, as the origin of the resistance oscillations. The present experiment, thus, indicates a feasibility of the significant re-population of Landau levels by the electric current.

\section{Acknowledgements}

S. V. thanks I. L. Aleiner for valuable help with the theoretical model and discussion. S. V. thanks Science Division of CCNY, CUNY Office for Research, PSC-CUNY Research Award Program (Project No. 63413-0041) and NSF (Grant No. DMR 1104503) for support of the experiments. A. A. B. thanks the Russian Foundation for Basic Research, Project No. 11-02-00925.

\end{document}